\documentclass[english,aip,jcp,reprint]{revtex4-1}
\usepackage[latin9]{inputenc}
\usepackage{babel}
\usepackage{amsmath}
\usepackage{amssymb}
\usepackage{graphicx}
\usepackage[unicode=true,
 bookmarks=false,
 breaklinks=false,pdfborder={0 0 1},colorlinks=false]
 {hyperref}
\usepackage{breakurl}

\makeatletter

\newcommand{\noun}[1]{\textsc{#1}}
\DeclareFontEncoding{LGR}{}{}
\DeclareRobustCommand{\greektext}{%
  \fontencoding{LGR}\selectfont\def\encodingdefault{LGR}}
\DeclareRobustCommand{\textgreek}[1]{\leavevmode{\greektext #1}}
\ProvideTextCommand{\~}{LGR}[1]{\char126#1}

 \@ifundefined{textcolor}{}
 {%
   \definecolor{BLACK}{gray}{0}
   \definecolor{WHITE}{gray}{1}
   \definecolor{RED}{rgb}{1,0,0}
   \definecolor{GREEN}{rgb}{0,1,0}
   \definecolor{BLUE}{rgb}{0,0,1}
   \definecolor{CYAN}{cmyk}{1,0,0,0}
   \definecolor{MAGENTA}{cmyk}{0,1,0,0}
   \definecolor{YELLOW}{cmyk}{0,0,1,0}
 }
\newcommand{\code}[1]{\texttt{#1}}

\usepackage{listings}
\usepackage{xfrac}

\usepackage{listings}

\usepackage{listings}

\makeatother

\usepackage{listings}

\begin{document}

\title{Graph-theoretical evaluation of the inelastic propensity rules for
molecules with destructive quantum interference }

\author{Rudolf Sýkora}

\affiliation{Department of Condensed Matter Physics, Faculty of Mathematics and
Physics, Charles University, Ke Karlovu 5, CZ-121 16 Praha 2, Czech
Republic}

\author{Tom\'a\v{s} Novotný}
\email{tno@karlov.mff.cuni.cz}

\selectlanguage{english}%

\affiliation{Department of Condensed Matter Physics, Faculty of Mathematics and
Physics, Charles University, Ke Karlovu 5, CZ-121 16 Praha 2, Czech
Republic}
\begin{abstract}
We present a method based on graph theory for evaluation of the inelastic
propensity rules for molecules exhibiting complete destructive quantum
interference in their elastic transmission. The method uses an extended
adjacency matrix corresponding to the structural graph of the molecule
for calculating the Green function between the sites with attached
electrodes and consequently states the corresponding conditions the
electron-vibration coupling matrix must meet for the observation of
an inelastic signal between the terminals. The method can be fully
automated and we provide a functional website running a code using
Wolfram Mathematica, which returns a graphical depiction of destructive
quantum interference configurations together with the associated inelastic
propensity rules for a wide class of molecules. 
\end{abstract}
\maketitle

\section{Introduction}

Transport through molecules exhibiting quantum interference features
due to multiple electronic paths connecting the two leads has been
in the past decade a subject of intensive research both theoretically\cite{Gemma:JACS08,Peskin:MolPhys08,Japonci:JACS08,Ratner:JChemPhys09,Markussen10,Troels:PCCP11,Gemma:Beilstein11,Hartle:PRL11,Lovey:ChemPhysLett12,Lambert:review15,Jens:PRB14,Lambert:JACS15,Pedersen15,Stadler:JChemPhys17,Paaske:preprint16,Reuter:preprint17}
and experimentally.\cite{Guedon:NatNano12,Latha:NatNano12,Weber:PRL12,Lambert:NatComm2015,Troels:SciRep16}
From the very beginning this subject has been addressed not only by
direct numerical approaches based on various levels of ab-initio calculations
but also by model studies, whose primary task is to bring conceptual
understanding of quantum interference effects.\cite{Lambert:review15}
The idea behind these attempts relies on the believed existence of
a connection between basic structural information on a molecule and,
at least qualitative, predictive power of some simple analytical procedure
(``rule'') indicating the (non)existence and potentially even magnitude
of quantum interference effects for the given molecule in a given
transport setup.

During the years there appeared a number of such rules formulated
either in terms of molecular-orbitals\cite{Gemma:JACS08,Japonci:JACS08,Lovey:ChemPhysLett12,Jens:PRB14}
or local tight-binding basis.\cite{Markussen10,Troels:PCCP11,Geerlings:JPCC15,Lambert:JACS15}
The implicit requirement of robustness and simplicity of such rules
basically forces them to rely on some very rudimentary properties
of the molecular structure, which are often of topological nature
and can be described and handled by graph theory. Indeed, such graph-theoretical
methods do actually have strong tradition in chemistry in various
contexts and flavors. \cite{Balaban,ChemGraphTh,Trinajstic} They
have been eventually applied also to the problem of quantum interference
in electronic transport through molecules. So far, however, their
application has been limited to the elastic transmission issue.

In this work, we formulate a graph-theoretical approach to the propensity
rules for inelastic electron tunneling spectroscopy (IETS) signals
for molecules which exhibit the destructive quantum interference (DQI)
in their elastic transmission, i.e., complete elastic current suppression.
Interestingly, we find out that the problem can be formulated equivalently
to the elastic case with a modified molecular Hamiltonian that includes
the electron-vibration coupling matrix. Therefore, all the previous
knowledge concerning the elastic case can be straightforwardly transferred
to the inelastic problem. Furthermore, to spare readers the necessity
of error-prone implementation of elastic graphical rules we present
a website which runs efficient and reliable Mathematica code evaluating
both the elastic transmission and inelastic IETS propensity rules,
using the Wolfram chemical database of molecules.

\section{Elastic transmission nodes from graph theory\label{sec:Elastic}}

The simplest quantum interference feature studied are the nodes in
the elastic transmission caused by the complete destructive quantum
interference. In this case, one is only after qualitative information
whether or not such a node will be present in the transmission function
close to the Fermi energy and, thus, achievable by applying a moderate
voltage bias. All the developed different approaches (graphical rules,\cite{Markussen10}
magic ratios,\cite{Lambert:JACS15} or the curly arrow\cite{Geerlings:JPCC15})
rely on the very same underlying approximation of the molecular Hamiltonian
by its Hückel form with one local orbital per atom and hopping connectivity
determined by the structural graph of the molecule. They subsequently
deal differently with this model Hamiltonian, however, they all use
the molecular structural graph's adjacency matrix as the basic Hückel
Hamiltonian. This is the core of the graph-theoretical approach that
we also adopt here.

It should be mentioned that this starting point is not free of questions.
In particular, it is obvious that the Hückel model is largely oversimplified
and unrealistic and, therefore, one must ask to what extent can the
conclusions drawn from it be taken seriously. A rather thorough comparison
of graphical-rules predictions and corresponding DFT calculations
for a number of nontrivial molecular structures in Ref.~\cite{Markussen10}
showed quite surprising one-to-one correspondence between the two
methods. The alleged breakdown of the graphical rules for azulene
in Ref.~\cite{Xia14} turned out to be caused by their incorrect
application (and was directly contradicted by the experimental data
within the same paper \cite{Xia14}) and, despite its weaknesses discussed
in the past year in the literature \cite{Stadler15Comm,Strange15Comm}
and also in our Appendix \ref{sec:MST10rules}, the graphical approach
seems to provide predictions on the presence/absence of the transmission
node with unrivaled accuracy.

Its stunning success can be partly understood by the molecular-orbital
point of view \cite{Lovey:ChemPhysLett12,Jens:PRB14} \textemdash{}
under reasonable assumptions the existence of a transmission node
within the HOMO-LUMO gap is determined exclusively by the signs of
the HOMO and LUMO molecular-orbital wavefunctions at the sites connected
to the leads. Such properties appear to be largely topological, i.e.,
independent of particular approximations and this explains the perfect
agreement between ab-initio DFT approach and simplistic Hückel method.
At the same time the exact energetic position of the transmission
node (which is always at zero energy for the Hückel model) depends
strongly on the used approximation, yet its existence within the HOMO-LUMO
gap is a topological property independent of the used method. More
elaborate many-body schemes such as GW may reorder the molecular orbitals
with respect to DFT and then the predictions on the (non)existence
of the transmission node within the HOMO-LUMO gap may differ.\cite{Jens:PRB14}
Nevertheless, this situation is relevant mostly for molecules weakly
coupled to the leads where correlation effects due to local Coulomb
interaction on the nearly isolated molecule are significant, which
is not the generically studied situation.\footnote{Very recently, two works \cite{Stadler:JChemPhys17,Reuter:preprint17}
have addressed in detail general questions concerning the DQI origin,
predictability and classification.}

\begin{figure}
\includegraphics{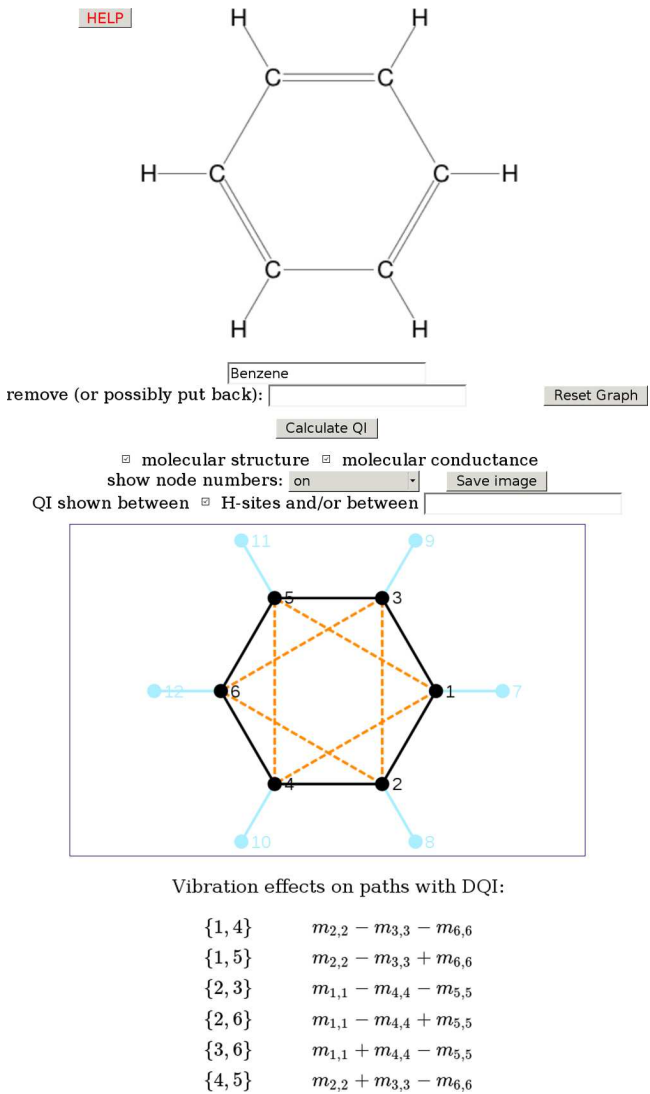} \caption{QI webpage with benzene as an example. For description see Sec.~\ref{sec:Elastic}
and especially App.~\ref{sec:User-manual}.\label{FIG:qiwww}}
\end{figure}

The correct formulation of the graphical rules requires for larger
molecules an elaborate enumeration and summation of graph diagrams\cite{Wai-KaiChen},
which is prone to error as the case of azulene clearly demonstrated.\cite{Xia14}
For this reason we adopted a safer computer-aided attitude to the
problem and developed a Mathematica code interfaced on a webpage \url{http://qi.karlov.mff.cuni.cz:1345}
which identifies the elastic DQI molecular configurations as well
as calculates the corresponding inelastic propensity rules for them.
A print-screen of the webpage is shown in Fig.~\ref{FIG:qiwww} for
the case of the benzene molecule. The user manual from the webpage
(red HELP button on the left top) is for convenience of the reader
reproduced in Appendix \ref{sec:User-manual} while technical details
about the service are given in Appendix \ref{sec:Code-Details}.

The code evaluates the elastic transmission using the well-known formula
\cite[Ch. 8]{Cuevas} 
\begin{equation}
T_{\mathrm{el}}(\varepsilon)=\mathrm{Tr}\left[\Gamma_{L}G(\varepsilon)\Gamma_{R}G^{\dagger}(\varepsilon)\right],\label{eq:transmission}
\end{equation}
with 
\[
G(\varepsilon)\equiv\left(\varepsilon-H_{\mathrm{mol}}+i\frac{\Gamma_{L}+\Gamma_{R}}{2}\right)^{-1},
\]
which under the assumption of sufficiently slow energy dependence
yields a constant differential conductance $dI/dV=2e^{2}/h\times T_{\mathrm{el}}(\varepsilon_{F})$.
Adopting the same approach as Markussen, Stadler, and Thygesen\cite{Markussen10}
(MST10), i.e.~using the Hückel model with one $p_{z}$ orbital per
atom and assuming that the leads couple to just single sites (atoms)
denoted $L/R$ (left/right) we have 
\begin{equation}
\Gamma_{L}=\gamma_{L}|L\rangle\langle L|,\quad\Gamma_{R}=\gamma_{R}|R\rangle\langle R|,\label{eq:locality}
\end{equation}
which then implies 
\begin{equation}
T_{\mathrm{el}}(\varepsilon_{F})=\gamma_{L}\gamma_{R}|\langle L|G(\varepsilon_{F})|R\rangle|^{2},\label{eq:elastic}
\end{equation}
and search for the transmission nodes reduces to the evaluation of
a specific element of the Green function at the Fermi energy which
is set to zero in the Hückel model, i.e.~$\varepsilon_{F}\equiv0$.
As explained in Ref.~\cite{Markussen10}, if the molecular Hamiltonian
$H_{\mathrm{mol}}$ is regular, i.e. $\det(H_{\mathrm{mol}})\neq0$,
the $LR$ matrix element of the full Green function (with the leads
included) is zero if and only if the corresponding element of the
molecular resolvent $G_{LR}^{0}\equiv\langle L|(\varepsilon_{F}-H_{\mathrm{mol}})^{-1}|R\rangle$
is also zero. Consequently, the existence or not of the DQI between
given pairs of atoms is determined exclusively by the molecular structure
itself and (within the employed approximations) does not depend on
the leads.

After we send a query by means of entering a molecule name, the server
responds with the molecule structural formula drawn on top of the
page together with its editable graph representation below, see Fig.~\ref{FIG:qiwww}.
Now, user must define the subgraph corresponding to the conjugated
backbone of the molecule (for details of this crucial operation consult
the instructions in the User manual in Appendix \ref{sec:User-manual};
sometimes the default processing of the molecule by the code without
user's intervention might be sufficient) and then can ask for the
calculation of the DQI pattern by pressing the \noun{Calculate QI}
pink button. The code takes the adjacency matrix of the selected subgraph,
which (up to a nonzero multiplicative factor) corresponds to the Hückel
model Hamiltonian and calculates its inverse (recall that $\varepsilon_{F}\equiv0$
in this model). Zero entries in this inverse, if they correspond to
a pair of vertices (atoms) which can be in principle contacted by
leads (typically atoms with hydrogens), are collected and depicted
in the graph by orange dashed lines (Fig.~\ref{FIG:qiwww}). These
are the configurations which exhibit full DQI in the \emph{elastic
regime}, i.e. for sufficiently small applied voltage below the excitation
threshold of molecular vibration modes. For \emph{these configurations}
the code also calculates the \emph{inelastic} contributions to the
conductance (i.e.\emph{~inelastic propensity rules}) as explained
in Sec.~\ref{sec:InelPropRules} which are displayed at the bottom of the webpage
under the text \noun{Vibration effects on paths with DQI:}

There are molecules whose graph representations yield singular matrices,
i.e.~their determinant is zero and there exist one or more eigenstates
exactly at zero energy. Examples of such a situation involve conjugated
linear chains with odd number of atoms (propene, pentadiene, etc.)
having a single zero mode and cyclobutadiene (corresponding to the
square graph, see Appendix \ref{sec:MST10rules} for more details
on this very interesting case) with two zero modes. In such cases
the above construction equating the full Green function and the molecular
resolvent does not hold and one has to use a more careful approach
involving the pseudoinverse of the molecular Hamiltonian described
in Appendix \ref{sec:Pseudoinverse}. Using this method our code still
does find the \emph{elastic} DQI configurations, but it does not calculate
the inelastic propensity rules since the extension of the inelastic
theory to this case has not been done yet.

\section{Inelastic propensity rules\label{sec:InelPropRules}}

\subsection{Derivation}

Now, we extend the graph-theoretical method to the evaluation of inelastic
signals. It was shown more than a decade ago \cite{Paulsson:RapCom,Viljas,delaVega}
that for weak electron-vibration coupling assumed here the inelastic
contribution of vibrational mode $\lambda$ to the differential conductance
takes on a generic form (Lowest Order Expansion, LOE - see Eqs.~(5)\textendash (7)
of Ref.~\cite{Paulsson:RapCom}) with two (symmetric and asymmetric)
universal functions of bias, mode frequency, and temperature multiplied
by system-dependent coefficients

\begin{multline}
\Delta_{\lambda}^{\mathrm{Sym}}=\mathrm{Tr}\Big[G^{\dagger}\Gamma_{L}G\big\{ M_{\lambda}G\Gamma_{R}G^{\dagger}M_{\lambda}+\\
+\frac{i}{2}(\Gamma_{R}G^{\dagger}M_{\lambda}AM_{\lambda}-\mathrm{h.c.})\big\}\Big].\\
\Delta_{\lambda}^{\mathrm{Asym}}=\mathrm{Tr}\Big[G^{\dagger}\Gamma_{L}G\big\{\Gamma_{R}G^{\dagger}M_{\lambda}G(\Gamma_{R}-\Gamma_{L})G^{\dagger}M_{\lambda}+\mathrm{h.c.\}}\Big],\label{eq:IETS}
\end{multline}
with the spectral function $A\equiv i(G-G^{\dagger})$ and electron-vibration
coupling matrix $M_{\lambda}$. All the Green functions are evaluated
at the Fermi energy $\varepsilon_{F}=0$ which corresponds to the
wide-band-limit (WBL) assumption of energy-independent $\Gamma$s
and $G$ used in the derivation of the above formulas \cite{Paulsson:RapCom}.
We will justify the validity of this assumption even in cases with
DQI shortly, but first let us simplify the above formulas for our
Hückel model with leads attached in a configuration exhibiting the
elastic DQI, i.e.~$\langle L|G(\varepsilon_{F})|R\rangle=0$. Using
this property together with Eq.~\eqref{eq:locality} in \eqref{eq:IETS}
we find easily that both terms in $\Delta_{\lambda}^{\mathrm{Asym}}$
as well as the two last terms (multiplied by $i/2$) in $\Delta_{\lambda}^{\mathrm{Sym}}$
are zero and we are only left with the first term

\begin{equation}
\Delta_{\lambda}^{\mathrm{Sym}}=\gamma_{L}\gamma_{R}\left|\langle L|GM_{\lambda}G|R\rangle\right|^{2}\approx\gamma_{L}\gamma_{R}\left|\langle L|G_{M_{\lambda}}|R\rangle\right|^{2},\label{eq:inelastic}
\end{equation}
where the second approximate equality holds in the lowest (second)
order in $M_{\lambda}$ and we have introduced a modified Green function
$G_{M_{\lambda}}$ that is obtained by substituting the molecular
Hamiltonian by its modification $H_{\mathrm{mol}}\rightarrow H_{\mathrm{mol}}+M_{\lambda}$.
By this step we have reformulated the inelastic problem equivalently
to the elastic problem, just with a modified molecular Hamiltonian,
see Eqs.~\eqref{eq:elastic} and \eqref{eq:transmission} above.
The modification involves the electron-vibration coupling matrix $M_{\lambda}$
specific for a given vibrational mode $\lambda$. To keep the Hückel-model
approach consistent, we only allow matrix elements of $M_{\lambda}$
to be nonzero on the structural graph of the molecule (more precisely,
of the molecular conjugated backbone), i.e.~just the diagonals and
off-diagonals corresponding to the nearest neighbors connected by
a chemical bond.

In principle, we can now adopt some form of the ``graphical rules''
for the evaluation of the elastic transmission mentioned in 
Sec.~\ref{sec:Elastic} to the calculation of the matrix element $\langle L|G_{M_{\lambda}}|R\rangle$
of the modified Green function. For this purpose, we use the very
same Mathematica-based code which now returns the matrix elements
of the modified Green function for all combinations of vertices exhibiting
DQI in the elastic transmission. We collect just the linear combinations
of the coupling matrix elements neglecting possible $\gamma$-dependent
but $M$-independent prefactor and display them in the MathML format
at the bottom of the webpage, cf.~Fig.~\ref{FIG:qiwww}. According
to Eq.~\eqref{eq:inelastic} their absolute value squared is proportional
to the intensity of the inelastic signal for the given vibronic mode,
which reveals itself as a jump in the differential conductance at
the vibrational excitation threshold $eV_{\mathrm{th}}=\hbar\omega_{\lambda}$.
Consequently, they constitute \emph{inelastic propensity rules} (in
the spirit of Eq.~(4) in Ref.~\cite{Paulsson:PRL08}) which can
be used for assessing the effects of molecular symmetries (and/or
other factors) for given vibrational modes. Contributions from different
vibrational modes are within LOE additive. 

\subsection{Justification and range of validity}

Now, let us discuss the applicability of Eqs.~\eqref{eq:IETS} to
the case with DQI in elastic transmission. As already mentioned the
microscopic derivation of these equations from the non-equilibrium
Green functions theory in Refs.~\cite{Paulsson:RapCom,Viljas,delaVega}
uses the WBL assumption of energy-independent $G(\varepsilon)$ in
the range of order $\hbar\omega_{\lambda}$ around the Fermi energy
$\varepsilon_{F}$. The validity of this assumption is certainly not
obvious in the DQI case because the Green function $G(\varepsilon)$
passes through zero at the energy of the antiresonance feature (in
the Hückel model at the Fermi energy $\varepsilon_{F}=0$) and is
finite around it. Thus it is certainly \emph{not} \emph{constant}
in the relevant energy range. However, it turns out that the width
of the elastic antiresonance is usually bigger than the typical vibronic
energy and, consequently, $G$ can be reasonably approximated by zero
in the whole range.

\begin{figure}
\includegraphics{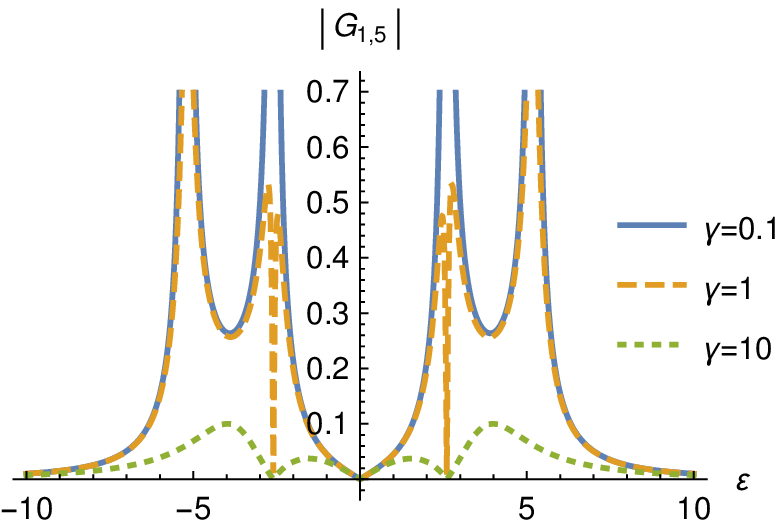} \\[.7cm] \includegraphics{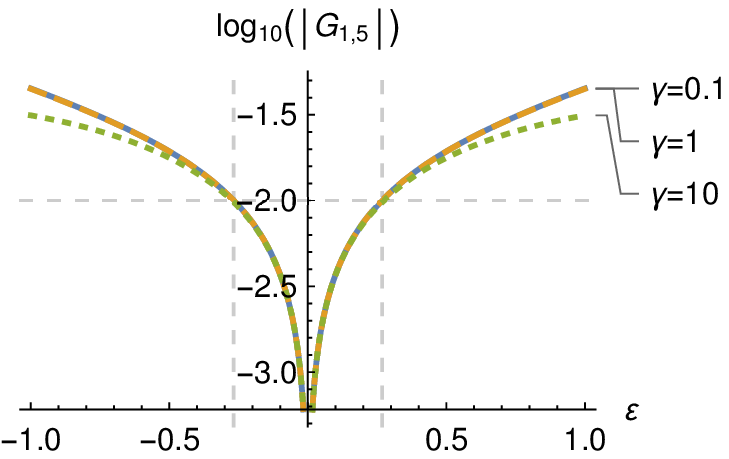} \caption{Green function of the meta-benzene for several values of the coupling
to the leads $\gamma$ (linear scale in the top panel and low-energy
zoom on the logarithmic scale in the bottom panel). We used Hückel
hopping element of 2.6\,eV according to Ref.~\cite[Sec. 9.5.1, p. 250]{Cuevas}.
Values of $\varepsilon$ and $\gamma$ are in eV, of $G$ in eV$^{-1}$.\label{fig:Benzene-elastic}}
\end{figure}

We illustrate this concept in Fig.~\ref{fig:Benzene-elastic} where
the magnitude of the Green function $|\langle L|G(\varepsilon)|R\rangle|$
for the benzene molecule contacted in the meta position ($L=1,\,R=5$)
exhibiting the DQI is plotted both on the linear (upper panel with
wider energy scale) as well as logarithmic (lower panel with zoom
on low-energy region) scale. The calculation uses the Hückel model
with the nearest-neighbor hopping amplitude $2.6$~eV \cite[Sec. 9.5.1, p. 250]{Cuevas}
and several values of $\gamma\equiv\gamma_{L}=\gamma_{R}$. While
on a larger energy scale above $1$ eV the Green function depends
strongly on the value of $\gamma$ (top panel), close to zero $|\varepsilon|\lesssim0.5$
eV the curves collapse (bottom panel) and if we choose our threshold
for ``machine zero'' as $0.01$ eV$^{-1}$ (gray dashed horizontal
line in the bottom panel corresponding to about 1\% of the maximal
values of the Green function for not too large $\gamma$s) we see
that the ``zero value'' of the Green function spans the range of
roughly $\pm300$ meV which matches the highest vibrational frequencies
of the conjugated backbone corresponding to stretching double and/or
triple C-C bonds.

Thus, this rough estimate shows that the usage of Eqs.~\eqref{eq:IETS}
is justified. Moreover, there exists an extension of Eqs.~(\ref{eq:IETS})
relaxing the WBL assumption \cite{Lu:PRB2014} \textemdash \ comparison
of the two approaches by ab-initio treatment \cite{Fabian} of the
meta-benzene and 3-methylene-1,4-pentadiyne molecules exhibiting DQI
(considered in Refs.~\cite{Lykkebo:ACSNano13,Lykkebo:JChemPhys14})
yields nearly identical results which further justifies the usage
of the WBL formula (\ref{eq:IETS}).

Concerning the range of validity of our approach, reader must be aware
of the basic fact that the methods works correctly for the \emph{Hückel
model} with equal hoppings between adjacent atoms (if connected by
the conjugated backbone). In reality, this assumption may not be realistic
and different values of hopping elements might be appropriate at certain
links due to, e.g.~non-planarity of the molecule (such as biphenyl
in Sec.~\ref{subsec:Biphenyl}) or Jahn-Teller distortion (as in
cyclobutadiene studied in App.~\ref{sec:MST10rules}). In such cases
one must be careful and explicitly address the effects of such a modification
of the molecular Hamiltonian on the DQI pattern. The results of our
code then serve only as the zeroth iteration of a more elaborate study
which may survive the refinement of the theory (as in the case of
biphenyl where the DQI pattern does not depend on the value of the
hopping element in the interlink between the benzene rings) or not
(see the Fano resonance analysis in cyclobutadiene at the end of App.~\ref{sec:MST10rules}).
Unfortunately, there is no a priori general prescription how to decide
whether our method is sufficient or not beyond its Hückel model paradigm
and one has to assess this case-by-case. 

Finally, as the Hückel model only deals with the $\pi$-system and
neglects the $\sigma$-system, it is obvious that some of the vibrational
modes (those coupled within the $\sigma$-system or coupling $\sigma$-system
to the $\pi$-system) are invisible to our approach. This is a shortcoming
of our approach and price to pay for the simplification of considering
only the $\pi$-system. The results for modes coupling within the
$\pi$-system should be, however, fully reliable and relevant. Thus,
we may say that our theory gives results only for a (important) subset
of vibrational modes, those coupled within the $\pi$-system, which
includes various C-C stretching modes etc. 

\subsection{Physical interpretation }

\begin{figure}
\includegraphics{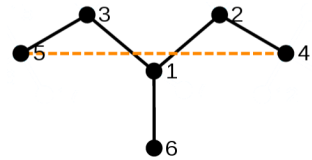} \caption{Graph representation of the 3-methylene-1,4-pentadiyne molecule. Only
4\textendash 5 DQI is shown, although it occurs also in other configurations.\label{fig:cross-conj}}
\end{figure}

We demonstrate the physical content of our procedure on a simplest
example of 3-methylene-1,4-pentadiyne molecule plotted in Fig.~\ref{fig:cross-conj}.
Unfortunately, this molecule is not known to the Wolfram chemical
database and, thus, our current form of the code cannot directly process
it. Nevertheless, we have used manual input into the Mathematica code
running at the background of the QI-webpage to obtain the results.
Moreover, as can be straightforwardly shown, here the results do \emph{not}
depend on the magnitude of the hopping element at the link $1-6$.
As shown in Fig.~\ref{fig:cross-conj} the end-to-end conductance
is suppressed due to DQI which is a consequence of a (broad) Fano
resonance analogous to the situation studied in App.~\ref{sec:MST10rules},
see Fig.~\ref{fig:square2chain}. The inelastic propensity rule for
this DQI configuration reads $m_{6,6}$, i.e.~the IETS signal is
nonzero if and only if the onsite electron-vibration matrix element
on the apex atom $6$ is non-zero. This can be easily understood in
terms of the which-path interferometer. The $4-5$ configuration DQI
in the (elastic) transmission is caused by destructive interference
of two paths \textemdash{} one is the direct one along the $4-2-1-3-5$
line while the other one is the indirect path $4-2-1-6-1-3-5$ digressing
to the side branch. At zero energy these two paths have the same amplitudes
and opposite phases and, therefore, cancel out exactly (full elastic
DQI).

Now, for sufficient applied voltage bias allowing excitation of a
vibrational mode the two paths may become distinguishable, which results
in lifting the DQI. This can happen if and only if a vibration is
excited at the apex atom $6$ \textemdash{} such a situation allows
for an identification of the used (indirect) path and DQI is thus
destroyed. On the other hand, vibrations with zero coupling element
at the apex atom $m_{6,6}=0$ in principle cannot distinguish the
two paths and, consequently, keep the DQI. Therefore, the inelastic
propensity rule is proportional solely to the element $m_{6,6}$.
We see that the physical principle behind the propensity rule in this
case is the simple which-path detection. For molecules with cycles
discussed in Sec.~\ref{sec:Results}, the results for the inelastic propensity
rules are considerably more complicated, yet, we do believe that the
basic physical mechanism behind them is analogous to the present case.
From the symmetry of the molecule, it is obvious that all vibrational
modes $\lambda^{a}$ antisymmetric with respect to the $1-6$ symmetry
axis of the molecule will necessarily have $m_{6,6}^{\lambda^{a}}=0$.
Therefore, the simple derived inelastic propensity rule reveals that
only symmetric vibrational modes can contribute to the IETS signal
of 3-methylene-1,4-pentadiyne.

\section{Results for selected molecules\label{sec:Results}}

In this section we apply the above-developed formalism to three molecules,
namely to benzene, biphenyl, and finally azulene. It should be stated
that our choice was somewhat random, selecting from some simple ``typical''
molecules encountered previously in the literature. Our code can easily
handle molecules of much higher complexity and we strongly encourage
the readers to experiment with molecule of their choice. Despite the
simplicity of our chosen molecules we still make interesting observations
concerning their inelastic propensity rules.

\subsection{Benzene}

The case of benzene is shown in Fig.~\ref{FIG:qiwww} in the form
of the output on our QI webpage resulting from the ``Benzene'' query.
We can see that the suppressed conductance configurations are just
those of the meta position of the leads. This has been well known
before. What is less obvious and known is the corresponding inelastic
propensity rule which nontrivially combines the diagonal electron-vibration
coupling elements on the adjacent sites to the leads. Closer inspection
of the rule reveals that the vibronic modes antisymmetric with respect
to the axis perpendicular to the connecting line of the leads (e.g.,
antisymmetric with respect to the $2-5$ axis for the $1-4$ lead
configuration) will nullify the expression by the symmetry. Correspondingly,
our theory predicts that only the symmetric modes may contribute to
the IETS signal. A more detailed quantitative study of this issue
will be presented elsewhere. \cite{Fabian}

\subsection{Biphenyl\label{subsec:Biphenyl}}

\begin{figure}
\includegraphics[width=8.5cm]{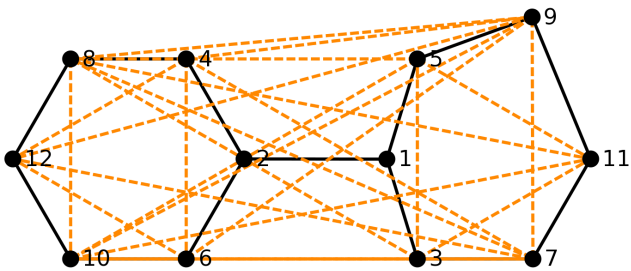}

\caption{Biphenyl; orange dashed lines connect configurations of leads (substitution
patterns) exhibiting elastic DQI. Vertices 1 and 9 are shifted to
reveal the otherwise overlapping DQI paths.\label{fig:Biphenyl}}
\end{figure}

Results for the biphenyl molecule are shown in Fig.~\ref{fig:Biphenyl},
where it is obvious that DQI pattern is rather complex for this more
complicated molecule (compared to the previous case of benzene). In
particular, the symmetry of the molecule may by overlapping lines
obscure which atoms are connected by the DQI lines and which are not.
To this end, the possibility of moving the graph vertices on the QI
webpage comes very handy and we show the resulting picture with the
vertices 1 and 9 moved away from their ``equilibrium'' positions
to reveal the full structure of the DQI network. On the IETS side
summarized in Eq.~(\ref{eq:Biphenyl}), we get specific combinations
of (\emph{only}) diagonal elements of the electron-vibration coupling
matrix. What we find interesting is the existence of two nonequivalent
DQI configurations ($7-8$ and $7-10$) for which there will be no
inelastic signal for all vibronic modes.

\begin{equation}
\begin{array}{ll}
\left\{ 3,5\right\}  & 4m_{1,1}+m_{4,4}+m_{6,6}-4m_{7,7}-4m_{9,9}+m_{12,12}\\
\left\{ 3,8\right\}  & 2\left(-m_{4,4}+m_{6,6}+m_{12,12}\right)\\
\left\{ 3,10\right\}  & 2\left(m_{4,4}-m_{6,6}+m_{12,12}\right)\\
\left\{ 3,11\right\}  & 4m_{1,1}+m_{4,4}+m_{6,6}-4m_{7,7}+4m_{9,9}+m_{12,12}\\
\left\{ 7,8\right\}  & 0\\
\left\{ 7,9\right\}  & 4\left(m_{3,3}+m_{5,5}-m_{11,11}\right)\\
\left\{ 7,10\right\}  & 0\\
\left\{ 7,12\right\}  & 2\left(m_{3,3}-m_{5,5}-m_{11,11}\right)
\end{array}\label{eq:Biphenyl}
\end{equation}

\subsection{Azulene}

\begin{figure}
\includegraphics[width=7cm]{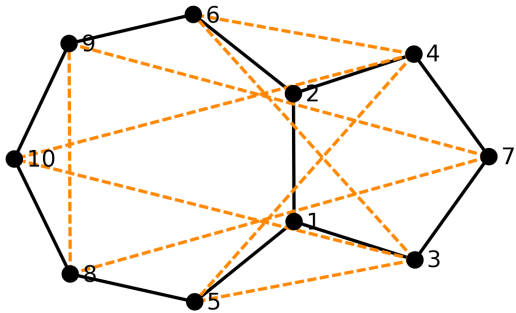}

\caption{Azulene; orange dashed lines connect configurations of leads (substitution
patterns) exhibiting elastic DQI..\label{fig:Azulene}}
\end{figure}

Study of an example of non-alternant hydrocarbon, azulene, was motivated
by the controversial work \cite{Xia14} and the results are given
in Fig.~\ref{fig:Azulene} and Eq.~(\ref{eq:Azulene}). We clearly
see that the disputed configuration $3-4$ correctly shows no DQI
while the other considered configuration $8-9$ does. We see interesting
features in the IETS propensity rules (\ref{eq:Azulene}) which, contrary
to the two previous cases of alternant hydrocarbons, contain the off-diagonal
elements of electron-vibration coupling matrix. We conjecture that
this property is specific of non-alternant hydrocarbons, which might
be worth further research efforts. 
\begin{widetext}
\begin{equation}
\mbox{\scriptsize\ensuremath{\begin{array}{ll}
\left\{ 3,5\right\}  & m_{1,1}+m_{1,2}-m_{1,5}-m_{2,1}-m_{2,2}+m_{2,6}+m_{3,1}-m_{3,7}-m_{4,2}+m_{4,7}-m_{7,7}+m_{8,5}-m_{8,8}-m_{8,10}-m_{9,6}-m_{9,9}+m_{9,10}\\
\left\{ 3,6\right\}  & m_{1,1}+m_{1,2}-m_{1,5}-m_{2,1}-m_{2,2}+m_{2,6}+m_{3,1}-m_{3,7}-m_{4,2}+m_{4,7}-m_{7,7}+m_{8,5}+m_{8,8}-m_{8,10}-m_{9,6}+m_{9,9}+m_{9,10}\\
\left\{ 3,10\right\}  & m_{1,1}+m_{1,2}-m_{1,5}-m_{2,1}-m_{2,2}+m_{2,6}+m_{3,1}-m_{3,7}-m_{4,2}+m_{4,7}-m_{7,7}+m_{8,5}+m_{8,8}-m_{8,10}-m_{9,6}-m_{9,9}+m_{9,10}\\
\left\{ 7,8\right\}  & m_{3,3}-m_{4,4}+m_{5,5}-m_{6,6}-m_{10,10}\\
\left\{ 8,9\right\}  & m_{3,3}+m_{4,4}+m_{5,5}+m_{6,6}-m_{10,10}
\end{array}}}\label{eq:Azulene}
\end{equation}
\end{widetext}

\section{Conclusions}

We have studied the influence of destructive quantum interference
effects in the elastic molecular transmission on the IETS signal via
a graph-theoretical approach. We have reformulated the inelastic propensity
rules in terms of the elastic problem with a modified molecular Hamiltonian,
which allows the application of graph-theoretical methods developed
for the elastic case also to the inelastic one. Moreover, we present
a Mathematica-based code for the calculation of the destructive quantum
interference configurations and inelastic propensity rules, which
can be accessed at the webpage \url{http://qi.karlov.mff.cuni.cz:1345}.
We have demonstrated our method on three simple example molecules
of benzene, biphenyl, and azulene, finding a rather complex structure
of DQI configurations and interesting features in the inelastic propensity
rules.

Our method is designed for a wide community of researchers who are
encouraged to play with the developed script and explore its results
for new molecules. As an open issue we leave the question of the universality
of graph-theoretical results for the inelastic quantities \textemdash{}
in particular, whether its prediction are as robust as for the existence
of the elastic transmission nodes. Preliminary ab-initio DFT results
for the IETS in meta-benzene \cite{Fabian} confirm our prediction
for the absence of the asymmetric vibrational modes stemming from
the symmetry analysis of the propensity rule, but more checks are
certainly needed to validate the method. Another interesting and relevant
aspect of the inelastic problem lies in the question whether there
is a simple method for assessing the electron-vibration coupling matrix
elements not requiring ab-initio calculations as an input. It would
be very helpful to have some version of ``inelastic Hückel model''
which would allow fully analytical study of the inelastic propensity
rules. As of today, we are not aware of any such model for the coupling
matrix and, in fact, even various ab-initio packages appear to give
vastly different results, as we discuss in our upcoming work \cite{Fabian}.
Thus this issue calls for attention of the IETS research community. 
\begin{acknowledgments}
We thank Jana Kalbá\v{c}ová Vejpravová for useful discussions on chemical
aspects of the present study. This work was supported by the Czech
Science Foundation by the grant No.\ 16-19640S. 
\end{acknowledgments}

\appendix

\section{Pseudoinverse\label{sec:Pseudoinverse}}

If we want to include molecules whose $H_{{\rm mol}}$ has a zero
determinant (i.e., $H_{{\rm mol}}$ has a zero eigenvalue), we may
proceed in several ways. As shown on the square graphs in Fig.\,\ref{FIG:squares}
in App.~\ref{sec:MST10rules}, one must be careful then. One could,
for each leads configuration, add non-zero selfenergies (due to leads
as well as infinitesimal $i\eta$s) at relevant places and
then calculate the inverse\textemdash however, this would be computationally
unnecessarily time-consuming. Below we present an alternative way,
which is used in our code.

Let $P$ be the projector on the null space of the isolated-molecule
Hamiltonian $H_{{\rm mol}}$, i.e., $H_{{\rm mol}}P=PH_{{\rm mol}}=0$,
and let $Q=1-P$. Then $PQ=QP=0$. The Green function of an isolated
molecule can then be rewritten as 
\begin{equation}
\begin{split}G_{0} & =\frac{1}{\varepsilon-H_{{\rm mol}}}=\frac{1}{\varepsilon(P+Q)-QH_{{\rm mol}}Q}=\\
 & =\frac{1}{P\varepsilon P+Q(\varepsilon-H_{{\rm mol}})Q}=\frac{P}{\varepsilon}+Q\frac{1}{\varepsilon-H_{{\rm mol}}}Q,
\end{split}
\end{equation}
valid whenever $(\varepsilon-H_{{\rm mol}})^{-1}$ exists. One infers
that in the interesting $\varepsilon\approx0$ region $G_{0}$ behaves
as 
\begin{equation}
\begin{split}G_{0} & =\frac{P}{\varepsilon}+\sideset{}{'}\sum_{i}\frac{|i\rangle\langle i|}{\varepsilon-\varepsilon_{i}}=\frac{P}{\varepsilon}-\sideset{}{'}\sum_{i}\frac{|i\rangle\langle i|}{\varepsilon_{i}}\big(1+\mathcal{O}(\frac{\varepsilon}{\varepsilon_{i}})\big)=\\
 & =\frac{P}{\varepsilon}-R(H_{{\rm mol}})+A\varepsilon+\mathcal{O}(\varepsilon^{2}),
\end{split}
\label{eq:G0expand}
\end{equation}
where in the $i$-summation over eigenvectors of $H_{{\rm mol}}$
we omit the terms corresponding to the zero eigenvalue, $R$ denotes
the Moore\textendash Penrose pseudoinverse, and $A$ is $\varepsilon$-independent.
Note that $P$ and $R$ are generally hermitian matrices. In our case
they are even real symmetric, since here-considered $H_{{\rm mol}}$
is real and symmetric; however, our formulae below do not rely on
this special property.

When leads are attached, the system's Green function is given by $G^{-1}=G_{0}^{-1}-\Sigma$.
Due to the assumed localization of $\Sigma$, the $LL$, $LR$, $RL$,
$RR$ elements of (retarded) $G$ are simply related in a $2\times2$
formalism thus 
\begin{equation}
\left(\begin{array}{cc}
G_{LL} & G_{LR}\\
G_{RL} & G_{RR}
\end{array}\right)^{-1}=\left(\begin{array}{cc}
G_{LL}^{0} & G_{LR}^{0}\\
G_{RL}^{0} & G_{RR}^{0}
\end{array}\right)^{-1}+\frac{i\gamma}{2}\left(\begin{array}{cc}
1 & 0\\
0 & 1
\end{array}\right).\label{eq:G2x2}
\end{equation}
To calculate conductance we need $G_{LR}$, which can be expressed
explicitly as 
\begin{widetext}
\begin{multline}
G_{LR}=4\varepsilon\big(P_{LR}-\varepsilon R_{LR}+\mathcal{O}(\varepsilon^{2})\big)\Big/\Big[\gamma^{2}(P_{LR}P_{RL}-P_{LL}P_{RR})+{}\\
+\varepsilon\Big(2i\gamma(P_{LL}+P_{RR})+\gamma^{2}\big(P_{LL}R_{RR}-P_{LR}R_{RL}-P_{RL}R_{LR}+P_{RR}R_{LL}\big)\Big)+{}\\
+\varepsilon^{2}\Big(4-2i\gamma(R_{LL}+R_{RR})-\gamma^{2}\big(R_{LL}R_{RR}-R_{LR}R_{RL}+A_{LL}P_{RR}-A_{LR}P_{RL}-A_{RL}P_{LR}+A_{RR}P_{LL}\big)\Big)+\mathcal{O}(\varepsilon^{3})\Big].\label{eq:GLR}
\end{multline}
\end{widetext}

We see that $G_{LR}$ may stay finite around $\varepsilon\approx0$
only if $P_{LR}P_{RL}-P_{LL}P_{RR}=0$. This is always so if $H_{{\rm mol}}$
does not have a zero eigenvalue ($P=0$), or even when such an eigenvalue
exists but is nondegenerate (labelling the corresponding normalized
vector $|0\rangle$ gives $P=|0\rangle\langle0|,$ from which the
result immediately follows). For a degenerate zero eigenvalue the
combination may or may not be zero. Nonetheless, from the Cauchy\textendash Schwarz
inequality $|\langle R|P|L\rangle|^{2}=|\langle R|PP|L\rangle|^{2}\leq\langle L|P|L\rangle\,\langle R|P|R\rangle$
it follows that whenever $P_{LL}$ or $P_{RR}$ is zero, so is $P_{LR}$
(and $P_{RL}$).

If $H_{{\rm mol}}$ does not have a zero eigenvalue ($P=0$), the
formula (\ref{eq:GLR}) in the $\varepsilon\rightarrow0$ limit simplifies
to 
\begin{equation}
G_{LR}^{'}=\frac{-4R_{LR}}{4-2i\gamma(R_{LL}+R_{RR})-\gamma^{2}(R_{LL}R_{RR}-|R_{LR}|^{2})}.\label{eq:GLRnoZero}
\end{equation}
We could remove $\mathcal{O}$s since the retained denominator cannot
become zero: both $R_{LL}$ and $R_{RR}$ would have to be zero to
nullify the imaginary part, but then we would be left with $4+\gamma^{2}|R_{LR}|^{2}>0.$
Hence $G_{LR}=0$ just when $R_{LR}=0$, as expected.

If a non-degenerate zero eigenvalue exists, then either (i) both $P_{LL}$
and $P_{RR}$ are zero, then $P_{LR}=P_{RL}=0$, and we once again
arrive at Eq.~(\ref{eq:GLRnoZero}), or (ii) at least one of $P_{LL}$,
$P_{RR}$ is nonzero (thus positive), the $\varepsilon$-term in the
denominator of (\ref{eq:GLR}) is necessarily nonzero due to the (single)
purely imaginary contribution, and we may disregard the higher-order
terms to obtain in the $\varepsilon\rightarrow0$ limit 
\begin{multline}
G_{LR}^{''}=4P_{LR}\Big/\Big[2i\gamma(P_{LL}+P_{RR})+{}\\
+\gamma^{2}\big(P_{LL}R_{RR}-2\Re(P_{LR}R_{RL})+P_{RR}R_{LL}\big)\Big].\label{eq:GLRdeg}
\end{multline}

Finally, if $H_{{\rm mol}}$ has a degenerate zero eigenvalue, either
$|P_{LR}|^{2}-P_{LL}P_{RR}\neq0$, in which case $G_{LR}=0$, or else
the situation is identical to the previous case of nondegenerate zero,
Eq.~(\ref{eq:GLRdeg}).

In any case, we see that from (\ref{eq:GLR}) we may\textemdash with
no assumptions\textemdash always remove the $\mathcal{O}(\varepsilon^{2})$
term from the numerator, the $\mathcal{O}(\varepsilon^{3})$ term
from the denominator, as well as all the terms containing $A$. In
the $\varepsilon\rightarrow0$ limit we may therefore simplify (\ref{eq:GLR})
to 
\begin{widetext}
\begin{multline}
G_{LR}=\lim_{\varepsilon\rightarrow0}4\varepsilon\big(P_{LR}-\varepsilon R_{LR}\big)\Big/\Big[\gamma^{2}(|P_{LR}|^{2}-P_{LL}P_{RR})+{}\\
+\varepsilon\Big(2i\gamma(P_{LL}+P_{RR})+\gamma^{2}\big(P_{LL}R_{RR}-2\Re(P_{LR}R_{RL})+P_{RR}R_{LL}\big)\Big)+{}\\
+\varepsilon^{2}\Big(4-2i\gamma(R_{LL}+R_{RR})-\gamma^{2}(R_{LL}R_{RR}-|R_{LR}|^{2})\Big)\Big].\label{eq:GLRused}
\end{multline}
\end{widetext}

This is the formula used in our program to calculate $G_{LR}$. We
stress once more that $P$ and $R$ are characteristics of the isolated
molecule ($H_{{\rm mol}}$) and do \emph{not} depend on leads positions.
Although we calculate them only once, equation (\ref{eq:GLRused})
provides $G_{LR}$ for any leads attachment.

\section{Comments on MST10 rules\label{sec:MST10rules}}

We have three points to make about the rules proposed in Ref.~\cite{Markussen10},
henceforth called the MST10 rules. We only discuss the underlying
graph theory and leave aside any practical (im)possibility to chemically
realize our graphs.\footnote{For example, we ignore the chemical instability of the cyclobutadiene
represented by the square graph in Fig.~\ref{FIG:squares}, which
we use just for illustration of our points.}

\subsubsection*{Sufficiency of $\boldsymbol{\det_{LR}(H_{\mathrm{mol}})=0}$}

\begin{figure}
\includegraphics[width=4cm]{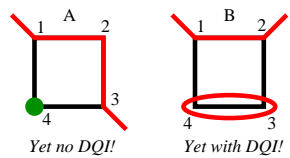}

\caption{Two graphs for which application of MST10 rules gives wrong prediction. }

\label{FIG:squares} 
\end{figure}

According to Ref.~\cite{Markussen10}, the condition for destructive
quantum interference (at $\varepsilon=0$) between molecular sites
$L$ and $R$ is given by their Eq.~(4), $\det_{LR}(H_{\mathrm{mol}})=0$.
However, the latter equation only follows from their equation Eq.~(3)
(which is our Eq.~(\ref{EQ:G1N}) below with the $i\eta$ terms removed)
if the denominator of their Eq.~(3) does not vanish. If it does vanish,
one ought to be more careful. An example of where this happens is
graph A in our Fig.\ \ref{FIG:squares}.

MST10 derive transmission through a molecule from their Eq.~(2),
$T(\varepsilon)=\gamma^{2}\left|G_{LR}(\varepsilon)\right|^{2}$ (the
same $\gamma$ is used for both leads). Although not stated there,
$G_{LR}$ to be considered is either the retarded or the advanced
limit (whichever) of the Green function. Hence, if we opt for the
latter, MST10's Eq.~(3) should read (for $\varepsilon=0$) in detail
\begin{equation}
G_{LR}=\lim_{\eta\rightarrow0^{+}}\frac{(-1)^{L+R}\det_{LR}(H_{\mathrm{mol}}+i\eta)}{\det(H_{\mathrm{mol}}+\Sigma_{L}+\Sigma_{R}+i\eta)},\label{EQ:G1N}
\end{equation}
where $\Sigma_{L}$ and $\Sigma_{R}$ are finite advanced self-energies
on sites $L$ and $R$, respectively, due to the attached leads; $\det_{LR}(A)$
is the determinant of $A$ from which the $L^{\mathrm{st}}$ row and
$R^{\mathrm{th}}$ column were removed. During the $\eta$-limiting
process both the numerator and the denominator may approach zero,
yet their ratio can stay finite.

For the graphs in Fig.\ \ref{FIG:squares} $H_{\mathrm{mol}}=\left(\begin{smallmatrix}0 & 1 & 0 & 1\\
1 & 0 & 1 & 0\\
0 & 1 & 0 & 1\\
1 & 0 & 1 & 0
\end{smallmatrix}\right)$ (we set our energy unit accordingly). For simplicity we take both
nonzero elements of $\Sigma_{L}$ and $\Sigma_{R}$ equal to $\sigma$.
In the case of graph A then 
\begin{equation}
G_{13}^{\mathrm{A}}=\lim_{\eta\rightarrow0^{+}}\frac{2i\eta}{4\eta^{2}+\eta^{4}-4i\eta\sigma-2i\eta^{3}\sigma+i\gamma^{2}\eta^{2}}=-\frac{1}{2\sigma}\label{eq:GA13}
\end{equation}
and stays finite. Especially, if we set $\sigma=i\gamma/2$, $\gamma\in\mathbb{R}$,
then $G_{13}^{\mathrm{A}}=i/\gamma$ and we arrive at a unit transmission,
$T=1$. We conclude that no destructive interference appears, contrary
to the MST10 prediction. Moreover, the result sounds natural, as we
have two exactly equivalent paths through the molecule and thus no
reason for any destructive interference.

The MST10 rule is correct in specifying when $\det_{LR}(H_{\mathrm{mol}})$
is zero; even in the case above. However, such a zero is not always
sufficient to imply zero conductance. Interestingly, the possibility
of having $\det(H_{\mathrm{mol}})$ zero is mentioned in Ref.~\cite{Stadler04}
(after their Eq.~(4)), but it was not elaborated on in any way there.
Also, Ref.~\cite{Pedersen15} touches the point in its supplement
after their Eq.~(3), but only vaguely, and referring to the strength
of coupling to the leads, which actually does not play any role in
our given example, since it cancels out in the resulting transmission.

\subsubsection*{Implication vs.~equivalence in the MST10 rules}

Although the impossibility to draw lines according to the MST10 rules
implies that $\det_{LR}(H_{\mathrm{mol}})$ is zero (simply because
this impossibility means that all the terms in the determinant expansion
are zero), one should not\textemdash contrary to what the other MST10
rule says\textemdash expect that a possibility to draw \emph{one}
MST10-rules-complying diagram automatically leads to a nonzero $\det_{LR}(H_{\mathrm{mol}})$.
This is exemplified by graph B in Fig.\ \ref{FIG:squares}. Here
\[
G_{12}^{\mathrm{B}}=\lim_{\eta\rightarrow0^{+}}\frac{\eta^{2}}{4\eta^{2}+\eta^{4}-4i\eta\sigma-2i\eta^{3}\sigma-\sigma^{2}-\eta^{2}\sigma^{2}}=0,
\]
the denominator does not vanish and only the numerator, i.e., $\det_{LR}(H_{\mathrm{mol}})$
plays a significant role. Thus we could actually avoid the use of
$\eta$, just as the original MST10 procedure would do. Anyway, we
obtain zero, while the MST10 rules predict it nonzero.

The problem is that the ability to draw a graph complying to the MST10
rules only means that there is a corresponding nonzero term in the
$\det_{LR}(H_{\mathrm{mol}})$ expansion, not that the sum of all
such terms with possibly varying signs is nonzero. Drawing a conclusion
from just one graph like B in Fig.\ \ref{FIG:squares} is erroneous,
all contributing graphs must be taken into account.

In this respect, it should be clearly stated that the relevant MST10
rule reading ``If such a continuous path can be drawn, then the condition
(4) is not fulfilled and a transmission antiresonance does not occur
at the Fermi energy.'' is incorrect. By omitting this fact, the comment
\cite{Stadler15Comm} on the rules'-breakdown article \cite{Xia14}
only prolonged the misunderstanding of some, which shows itself in
the reaction to the comment, \cite{Strange15Comm} stumbling on this
very point. Hopefully though, the necessity to consider all the graphs
seems now to have been (re-)established \cite{Troels:PCCP11,Pedersen15};
we note that the old paper \cite{Stadler04} actually did use the
summation.

\subsubsection*{Fano resonance in the square}

Above in this appendix we discussed conductance of a square with leads
attached to the opposite corners, Fig.~\ref{FIG:squares}A, and we
saw the conductance (at zero energy and for a symmetric leads attachment)
is equal to unity (see below Eq.~(\ref{eq:GA13})). While this is
true, there is an interesting extra point worth mentioning: if one
tries and calculates conductance of a distorted square in which not
all the hopping elements are the same, one finds that even an infinitesimal
distortion leads to an abrupt decrease of the conductance from one
to zero, a fact that perhaps deserves a comment.

First, the fact that we need $i\eta$ (and not just a finite $i\gamma$)
to ascertain that the inverse of $H_{\mathrm{mol}}$ exists suggests
that there is a completely decoupled, i.e., for transport irrelevant,
molecular eigenstate at zero energy. This is confirmed explicitly
after changing the basis from states $\left\{ \left|1\right\rangle ,\left|2\right\rangle ,\left|3\right\rangle ,\left|4\right\rangle \right\} $
(localized at the individual sites) to $\left\{ \left|1\right\rangle ,\left|+\right\rangle ,\left|3\right\rangle ,\left|-\right\rangle \right\} $
with $\left|+\right\rangle \equiv1/\sqrt{2}\big(|2\rangle+|4\rangle\big)$
and $\left|-\right\rangle \equiv1/\sqrt{2}\big(|2\rangle-|4\rangle\big)$.
In this basis $H_{\mathrm{mol}}^{'}=\sqrt{2}\left(\begin{smallmatrix}0 & 1 & 0 & 0\\
1 & 0 & 1 & 0\\
0 & 1 & 0 & 0\\
0 & 0 & 0 & 0
\end{smallmatrix}\right)$. We see that the system is actually equivalent to a 3-site linear
chain with all the hoppings equal to $\sqrt{2}$ and zero on-site
energies plus a completely isolated state. The chain (as any similar
chain with \emph{odd} number of sites) has unity transmission (at
zero energy) when connected to leads at its ends, here sites 1 and
3. This is the result we expect. 
\begin{figure}
\includegraphics{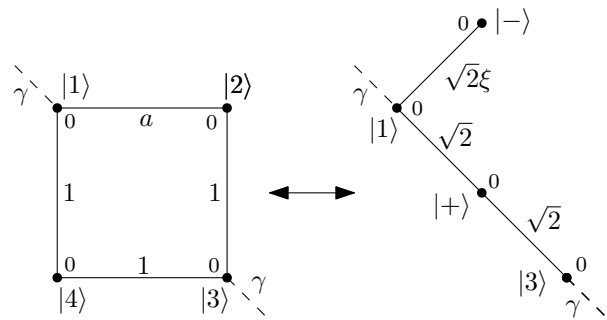}

\caption{Equivalence of an $a111$-disturbed square and a chain with a branch.}

\label{fig:square2chain} 
\end{figure}

\begin{figure}
\includegraphics{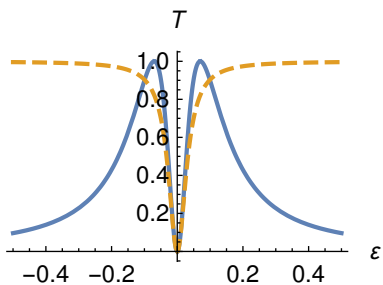}

\caption{Fano resonance in transmission $T$ through an $a111$-deformed square,
see Fig.\ \ref{fig:square2chain}, with $\gamma=0.3$ and $a=1.07$.
Blue (full) line is the full $\varepsilon$-dependence while orange
(dashed) line is its small-$\varepsilon$ approximation (\ref{eq:TSq12}).}

\label{fig:Fano} 
\end{figure}

Second, if we disturb one hopping from 1 to $a$, say, between sites
1 and 2\textemdash we call the disturbance $a111$\textemdash , and
again do the transformation, see Fig.\ \ref{fig:square2chain}, we
obtain $H_{\mathrm{mol}}^{'a111}=\frac{\sqrt{2}}{2}\left(\begin{smallmatrix}0 & a+1 & 0 & a-1\\
a+1 & 0 & 2 & 0\\
0 & 2 & 0 & 0\\
a-1 & 0 & 0 & 0
\end{smallmatrix}\right)\approx\sqrt{2}\left(\begin{smallmatrix}0 & 1 & 0 & \xi\\
1 & 0 & 1 & 0\\
0 & 1 & 0 & 0\\
\xi & 0 & 0 & 0
\end{smallmatrix}\right)$, where $\xi\equiv(a-1)/2\ll1$. This means that what before was an
isolated state now weakly couples to the chain. And this is the archetypal
characteristics of systems displaying the Fano resonance, see \cite[Sec. 13.6]{Cuevas}.
The formula (13.17) of the reference describes transport through a
single site (i.e., a 1-site chain) coupled to leads as well as to
a single other site with potentially different on-site energy. Our
current case is similar, though on-site energies are all zero and
we have a 3-site chain. Explicit calculation gives in the small-$\varepsilon$
limit and for $\xi/\gamma\ll1$ transmission 
\begin{equation}
T(\varepsilon\approx0)=\frac{1}{1+\big(\frac{2\xi^{2}}{\varepsilon\gamma}\big)^{2}}=\frac{1}{1+\big(\frac{(a-1)^{2}}{2\varepsilon\gamma}\big)^{2}},\label{eq:TSq12}
\end{equation}
which is essentially the reference's equation (with $\epsilon=\epsilon_{0}=0$,
$t=\sqrt{2}\xi$). We see our system displays a coexistence of two
processes: (i) in energy smoothly behaving transport along the chain
and (ii) a sharp (if $a$ is close to one) Fano resonance positioned
at $\varepsilon=0$, i.e., at the very place where the former smooth
part would otherwise have a maximum. Fig.~\ref{fig:Fano} depicts
this situation.\footnote{Since Fano resonances often occur to a side from other extrema, they
often have an asymmetric-in-energy signature. Not so it happens, however,
in our case.} 

If both the 1\textendash 2 and 3\textendash 4 hoppings were changed
to $a$ (disturbance $a1a1$), the state $|-\rangle$ would (after
the transformation) connect equally well to the $|1\rangle$ and $|3\rangle$
states, with the strength of $\sqrt{2}\xi$, and the transmission
would again include the Fano resonance (the denominator in Eq.~(\ref{eq:TSq12})
would be $1+(\frac{2(2\xi)^{2}}{\varepsilon\gamma})^{2}$).

In contrast, changing hoppings 1\textendash 2 and 4\textendash 1 to
$a$ (i.e., in a `symmetric' way; disturbance $a11a$) keeps the $|-\rangle$
state perfectly decoupled, irrespective of $a$, and the transmission
remains unity with no Fano feature present\textemdash as if the square
were perfect.

\section{User manual\label{sec:User-manual}}

\subsubsection*{NB}

Check the tooltips shown when you position the mouse over the entry
fields. After modifying any field press Enter there to reconsider
the value; pink color indicates changed but not yet considered values.

\subsubsection*{Molecule entry}

Enter the molecule name and press Enter to fetch the molecular data.
Molecules that are `autocompleted' are arenes known to Wolfram Alpha
(as of 7/2016); however, any molecule known to Wolfram Alpha can actually
be entered. You can also use molecule's CAS, CID, and perhaps also
other shown numbers. (E.g., entering Azulene, CID9231, and CAS275-51-4
yields the same.) The request can take some time, usually about 15
seconds.

\subsubsection*{Preparation for conductance calculation }

Before calculating conductance it is the user's responsibility to
prepare the relevant graph skeleton that should be composed only of
the molecule's conjugated (\textgreek{p}-bonded) subsystem(s). (Hence,
one generally removes at least the hydrogen atoms.)

Atoms listed in the `remove (or possibly put back):' entry are (if
existent) removed or reclaimed upon pressing Enter. Operations are
carried out in the written order, a leading `+' indicates reclamation,
otherwise removal takes place. Atoms are removed together with all
attached edges. Atoms can be designated by their label (`5'), by their
type (`H'), or by their type augmented by the number of attached edges
(`C:4'). After fetching molecular data, we by default remove H and
C:4 from the graph, since this is what one naturally almost always
wants. However, one can explicitly bring the atoms and related edges
back, using the provided tools, or more straightforwardly by pressing
the `Reset Graph' button. Both atoms and edges can be removed and/or
reclaimed by right-clicking them. Removal of an atom removes also
the attached edges, removal of an edge removes only the edge. Reclamation
of an atom reclaims only the atom, reclamation of an edge brings back
its end atoms as well. Atoms and edges removed from the molecular
graph are shown in light blue. Only objects in black enter conductance
calculation. The `remove (or possibly put back):' entry field does
not try to reflect all the changes made to the graph (e.g. with mouse
clicks). It only serves as a means of input for a single-shot operation.

\subsubsection*{Conductance and vibration effects calculation}

Calculation of conductance is started by clicking the `Calculate QI'
button. Dashed orange lines connect sites between which (total) destructive
quantum interference (DQI) takes place, effectively zeroing elastic
transmission. Note that such lines do not connect atoms not belonging
to the same conjugated part of the molecule (then, although we have
zero conductance, it is not due to interference, but due to effective
electrical isolation). The DQI (i.e., dashed orange) lines are only
shown between (i) atoms to which a hydrogen atom is attached in the
molecule, if the `H-sites' checkbox is checked, and/or (ii) any atoms
whose labels (numbers) have been inputted in the for-this-purpose
reserved field. If the considered-graph adjacency matrix has no inverse,
the user obtains a warning about zero-energy-mode(s) existence, along
with the zero-energy-eigenvalue degeneracy. For this case we still
provide the elastic conductance output, however, not the effect of
vibrations (the topic has not been fully investigated by us yet).
One should also be careful with the interpretation of the results
then, especially if several zero-energy modes exist, since molecules
in such cases rather undergo distortions (not accounted for here),
rendering provided predictions suspect
at least. More studies in this direction are needed. The `Vibration
Effects' table shows combinations of electron-vibration-interaction
(EVI) matrix elements, one for each currently displayed path featuring
DQI. An absolute-square of each such combination is the functional
dependence of the lowest-order change of the originally zero (due
to DQI) conductance on the EVI matrix elements. The `Vibration Effects'
table is in the MathML format, allowing reuse. {[}A right-click on
the table offers `Show Math As', `MathML Code'; this can be copied
to the clipboard (often ctrl-a ctrl-c) and inserted to a MathML-aware
application (such as LibreOffice: Insert - Object - Formula, Tools
- Import MathML from Clipboard). MathML can also be saved to a file
and/or converted to TeX, e.g. by an online tool here.{]}

\subsubsection*{Additional information}

Graph nodes can be moved with the mouse. The graph can be moved as
a whole by dragging and can be resized with the mouse wheel. The graph
can be, in its current state, saved by pressing the `Save image' button.
Pressing the `Reset Graph' button resets the controls and the graph
to the initial state, even before the default removal of H and C:4.
The checkboxes `molecular structure' and `molecular conductance' toggle
the visibility of molecular-structure and molecular-conductance edges,
respectively. The latter case is only applicable after conductance
has been calculated. The `show node numbers:' menu controls the visibility
of node labels (the numbers next to nodes). 

\section{Code details\label{sec:Code-Details}}

\subsubsection*{Service overview}

\begin{figure}
\includegraphics{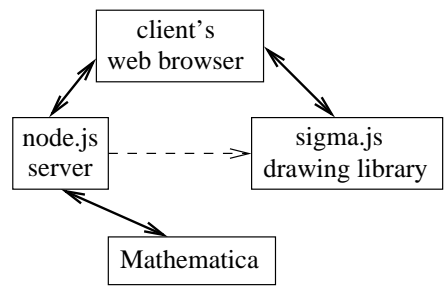} \caption{Program building blocks. \label{FIG:progDiag}}
\end{figure}

Our program is accessible as a web service at \url{http://qi.karlov.mff.cuni.cz:1345}.
Fig.\ \ref{FIG:qiwww} depicts a screenshot of the web page with
calculated conductance for benzene. To use the service, a web browser
with enabled javascript is necessary. The details about service usage
are provided under the Help button of the page. Fig.~\ref{FIG:progDiag}
shows the service building blocks together with their relations. In
order to provide a lucid representation of results, we use an interactive
graph-drawing (javascript) library sigma.js \cite{sigmajs}. Its built-in
interactivity allows easy rearrangement of graph layouts, which alleviates
the frequent problem of line overlaps. Sigma.js library gets downloaded
to the client's browser from our server upon directing the browser
to our service. The server runs the node.js framework \cite{nodejs}.

Input parameters for a calculation are taken from the web page and
sent via the server as parameters to Mathematica \cite{wolfram} scripts
that do the calculation. The results are, via standard output and
error streams, returned to the server, which relays them to the client's
browser for display.

The Mathematica component provides, besides the fairly simple computational
part described below, access to the Wolfram Research chemical database,
from which chemical structures of molecules are retrieved as graphs.
While this may limit the number of available molecules, this information
source relieves the burden of data entry. 

\subsubsection*{Mathematica scripts}

In the following, numbers enclosed in angle brackets denote lines
of the commented code.

The first Mathematica script is used to retrieve structural information
about the molecule entered on the web page. It reads an argument from
the command line to see what \code{molecule} is to be processed $\langle1\rangle$,
tries to fetch a graph of the molecule structure from the \code{ChemicalData}
database $\langle9\rangle$, and saves it to the variable \code{g}.
A picture of the graph in the Portable Network Graphics (PNG) format is sent Base64-encoded (since
we use text streams) to the output $\langle19\rangle$. This is the
picture one sees on the web page. Note that in the output stream we
delimit individual information pieces with tags, such as \code{M-PNG-Base64}
$\langle18,20\rangle$, used on the client's side to distinguish the
information. The rest of the script builds a JavaScript Object Notation (JSON) object populated
with all needed information about the molecule $\langle$22\textendash 34$\rangle$,
and outputs it $\langle$36\textendash 44$\rangle$. JSON objects
are easily handled by the sigma.js library, which draws the interactive
graph in the client's browser. 
\begin{lstlisting}[name=mathematicaScript]
material = $CommandLine[[-1]];
Print["MOLECULE"];
Print[material];
Print["/MOLECULE"];

(*get the data about the material*)
(*we do two checks since, for some reason, the response from
  Mathematica can vary*)
g = Check[ChemicalData[material, "StructureGraph"], 
  Write["stderr", "Molecule name not known."]; Quit[]];
If[MissingQ[g], Write["stderr", "Molecule name not known."];
  Quit[]];
(*seems easier to get VertexTypes from ChemicalData than
  from the graph g*)
gVTypes = ChemicalData[material, "VertexTypes"]

(*output PNG of the molecule graph*)
Print["M-PNG-Base64"];
Export[$Output, ExportString[g, "PNG"], "Base64"];
Print["/M-PNG-Base64"];

nodes = MapThread[{"id" -> ToString[#1],
  "label" -> ToString[#1],
  "x" -> #2[[1]], "y" -> -#2[[2]],
  "type" -> ToString[#3]} &,
  {VertexList[g], PropertyValue[g, VertexCoordinates], gVTypes}
  ];

nEdges = Length[EdgeList[g]];
edges = MapThread[Flatten[List[#1, #2]] &,
  {Thread["id" -> Map["e" <> ToString[#] &,
  Range[nEdges]]],
  ({"source" -> ToString[#1], "target" -> ToString[#2]} &)
    @@@ EdgeList[g]}];

Print["JSON"];
Check[
  Export[$Output,
  {"nodes" -> nodes,
  "edges" -> edges},
  "JSON"], Write["stderr", "Graph export failed."]; Quit[],
  Export::badval];
Print[];
Print["/JSON"];
\end{lstlisting}

After we determine, by removing vertices and/or edges from the original
graph, what the conjugated part of the molecule really is, we send
the modified graph to a second Mathematica script, which does the
calculation proper. Again, the graph is passed on the command line
(as a string) $\langle47\rangle$. For the graph we determine the
adjacency matrix \code{gAM} $\langle56\rangle$, and proceed along
the lines of Appendix \ref{sec:Pseudoinverse}, i.e., by calculating
the pseudoinverse \code{R} $\langle47\rangle$, projector \code{P}
$\langle$72\textendash 3$\rangle$, and $G_{LR}$ in the $\varepsilon\rightarrow0$
limit, in the code denoted by \code{GijLim} $\langle$75\textendash 83$\rangle$,
c.f. Eq.~(\ref{eq:GLRused}). Zero $G_{LR}$, if the sites $L$ and
$R$ are connected in the considered graph, means there is a total
destructive quantum interference (DQI) present, the fact which we
note down to \code{QI} $\langle$85\textendash 94$\rangle$ and output
as a list of edges where DQI occurs $\langle$100\textendash 2$\rangle$.

Finally, we calculate the effect of vibration modes on the conductance
of paths that without vibrations feature complete destructive interference.
It is (for the present) only calculated for the case when $H_{\mathrm{mol}}$
has no zero eigenvalue $\langle106\rangle$. As we discuss in the
main text, we introduce an electron-vibration coupling matrix \code{M}
having nonzero elements on its diagonal and between sites that are
directly connected in the graph $\langle$108\textendash 9$\rangle$,
and evaluate $G_{0}MG_{0}$ $\langle110\rangle$ (if $H_{\mathrm{mol}}$
has no zero eigenvalue, $G_{0}=-R$), absolute-square of which is
proportional to the lowest-order vibration-induced change in the conductance.
For each originally zero-conductance path we output the related combination
of matrix-$M$ elements (to be absolute-squared). We send such a table
in the MathML format to the client $\langle$134\textendash 6$\rangle$.

\begin{lstlisting}[name=mathematicaScript]
(*graph as a string like
 gStr = "{{n1,n2},{n2,n3},{n3,n4},{n4,n5},{n5,n6},{n6,n1}}";*)
gStr = $CommandLine[[-1]];

edges = UndirectedEdge[#[[1]], #[[2]]] & /@ ToExpression[gStr];
g = Graph[edges];
v = VertexList[g];

vLen = Length[v];
zM = ConstantArray[0, {vLen, vLen}];

gAM = AdjacencyMatrix[g];

R = PseudoInverse[gAM];

(*normed basis of nullspace of gAM*)
ns = (#/Norm[#] &) /@ NullSpace[gAM];
nsDeg = Length[ns];

Print["NS_DEGENERACY"];
Print[nsDeg];
Print["/NS_DEGENERACY"];

(*and corresponding projectors*)
PPerVector = Outer[Times, #, #] & /@ ns;

(*projector onto the whole nullspace*)
P = Apply[Plus, PPerVector];
If[P == 0, P = zM];

G[i_, j_] := (4 P[[i, j]] \[Omega] + 4 R[[i, j]] \[Omega]^2) /
  ((P[[i, j]]^2 - P[[i, i]] P[[j, j]]) \[Gamma]^2 +
   \[Gamma] (2 I P[[j, j]] - P[[j, j]] R[[i, i]] \[Gamma] + 
   2 P[[i, j]] R[[i, j]] \[Gamma] + P[[i, i]] (2 I - R[[j, j]] \[Gamma])) \[Omega] +
   (4 + 2 I R[[j, j]] \[Gamma] + R[[i, j]]^2 \[Gamma]^2 + 
   R[[i, i]] \[Gamma] (2 I - R[[j, j]] \[Gamma])) \[Omega]^2)

Gij = Table[G[i, j], {i, vLen}, {j, vLen}];
GijLim = Map[Limit[#, \[Omega] -> 0] &, Gij, {2}];

(*zero elements of GijLim are replaced with 1, nonzero with 0*)
QI = Map[If[# === 0, 1, 0] &, GijLim, {2}];

(*There is no real QI between disconnected parts*)
gDistM = GraphDistanceMatrix[g];
gConnM = Map[If[# == Infinity, 0, 1] &, gDistM, {2}];
QI = QI gConnM;

(*zero the diagonal*)
QI = ReplacePart[QI, {i_, i_} -> 0];

gQI = AdjacencyGraph[v, QI];
eQI = EdgeList[gQI];
eQIList = Apply[List, eQI, {1}];

Print["EQI"];
Print[eQIList];
Print["/EQI"];

(*Phonon part, now only when there is no zero energy
  in the spectrum*)
If[nsDeg == 0,
  (*create a matrix M that lives on g*)
  M = (Outer @@ {Subscript[m, #1, #2] &, #, #} & @ v)
    (gAM + IdentityMatrix[Length[v]]);
  GMG = R.M.R; (*NB we use the pseudoinverse*)
  
  pWith1 = Position[QI, 1]; (*where is 1 in QI?*)
  If[Length[pWith1] == 0, phononsEffect = "No path with DQI present",
    p = Pick[pWith1, (#[[1]] < #[[2]]&) /@ pWith1]; 
    
    GMGQI = GMG[[#[[1]], #[[2]]]] & /@ p; (*consider paths with QI*) 
  
    (*let's get rid of fractions*)
    Ph = GMGQI Det[gAM] // Simplify;
    (*just a simple check for a leading minus sign; can't make it
      worse, can it?*)
    Ph = If[Characters[ToString[#, InputForm]][[1]] == "-", -#, #]& /@ Ph;
    (*mapping between positions and vertices*)
    e = v[[#]]& /@ p;
  
    e = Sort /@ e;
    ePh = MapThread[List, {e, Ph}];
    ePh = Sort[ePh, #1[[1]] < #2[[1]] &];
  
    Mtable = TableForm[ePh, TableDepth -> 2];
    phononsEffect = ExportString[Mtable, "MathML"];
  ]

  Print["PHONONQI"];
  Print[phononsEffect];
  Print["/PHONONQI"];
]
\end{lstlisting}

\bibliography{QuantumInterference}

\end{document}